\documentclass[twocolumn,aps,prd,floatfix,nofootinbib,superscriptaddress]{revtex4-2}
\usepackage[utf8]{inputenc}
\usepackage{amssymb}
\usepackage{graphicx}
\usepackage{import}
\usepackage{mathtools}
\usepackage{hyperref}
\usepackage{mathrsfs}
\usepackage{multirow}
\usepackage{epstopdf}
\usepackage{longtable}

\begin{document}

\title{\boldmath Search for the process $\mathbf{\it e^+e^-\to\eta^{\prime}\gamma}$ in the energy range $\mathbf{\it \sqrt{s}}$ = 1.075 -- 2 GeV }

\author{M.~N.~Achasov}
\affiliation{Budker Institute of Nuclear Physics, SB RAS, Novosibirsk, 630090, Russia}
\affiliation{Novosibirsk State University, Novosibirsk, 630090, Russia}
\author{A.~Yu.~Barnyakov}
\affiliation{Budker Institute of Nuclear Physics, SB RAS, Novosibirsk, 630090, Russia}
\affiliation{Novosibirsk State University, Novosibirsk, 630090, Russia}
\affiliation{Novosibirsk State Technical University, Novosibirsk, 630073, Russia}
\author{E.~V.~Bedarev}
\affiliation{Budker Institute of Nuclear Physics, SB RAS, Novosibirsk, 630090, Russia}
\affiliation{Novosibirsk State University, Novosibirsk, 630090, Russia}
\author{K.~I.~Beloborodov}
\affiliation{Budker Institute of Nuclear Physics, SB RAS, Novosibirsk, 630090, Russia}
\affiliation{Novosibirsk State University, Novosibirsk, 630090, Russia}
\author{A.~V.~Berdyugin}
\email{A.V.Berdyugin@inp.nsk.su}
\affiliation{Budker Institute of Nuclear Physics, SB RAS, Novosibirsk, 630090, Russia}
\affiliation{Novosibirsk State University, Novosibirsk, 630090, Russia}
\author{A.~G.~Bogdanchikov}
\affiliation{Budker Institute of Nuclear Physics, SB RAS, Novosibirsk, 630090, Russia}
\author{A.~A.~Botov}
\affiliation{Budker Institute of Nuclear Physics, SB RAS, Novosibirsk, 630090, Russia}
\author{T.~V.~Dimova}
\affiliation{Budker Institute of Nuclear Physics, SB RAS, Novosibirsk, 630090, Russia}
\affiliation{Novosibirsk State University, Novosibirsk, 630090, Russia}
\author{V.~P.~Druzhinin}
\affiliation{Budker Institute of Nuclear Physics, SB RAS, Novosibirsk, 630090, Russia}
\affiliation{Novosibirsk State University, Novosibirsk, 630090, Russia}
\author{V.~N.~Zhabin}
\affiliation{Budker Institute of Nuclear Physics, SB RAS, Novosibirsk, 630090, Russia}
\affiliation{Novosibirsk State University, Novosibirsk, 630090, Russia}
\author{Yu.M.Zharinov}
\affiliation{Budker Institute of Nuclear Physics, SB RAS, Novosibirsk, 630090, Russia}
\author{L.~V.~Kardapoltsev}
\affiliation{Budker Institute of Nuclear Physics, SB RAS, Novosibirsk, 630090, Russia}
\affiliation{Novosibirsk State University, Novosibirsk, 630090, Russia}
\author{A.~S.~Kasaev}
\affiliation{Budker Institute of Nuclear Physics, SB RAS, Novosibirsk, 630090, Russia}
\author{A.~A.~Kattsin}
\affiliation{Budker Institute of Nuclear Physics, SB RAS, Novosibirsk, 630090, Russia}
\author{A.~N.~Kyrpotin}
\affiliation{Budker Institute of Nuclear Physics, SB RAS, Novosibirsk, 630090, Russia}
\author{D.~P.~Kovrizhin}
\affiliation{Budker Institute of Nuclear Physics, SB RAS, Novosibirsk, 630090, Russia}
\author{I.~A.~Koop}
\affiliation{Budker Institute of Nuclear Physics, SB RAS, Novosibirsk, 630090, Russia}
\affiliation{Novosibirsk State University, Novosibirsk, 630090, Russia}
\author{A.~A.~Korol}
\affiliation{Budker Institute of Nuclear Physics, SB RAS, Novosibirsk, 630090, Russia}
\affiliation{Novosibirsk State University, Novosibirsk, 630090, Russia}
\author{A.~S.~Kupich}
\affiliation{Budker Institute of Nuclear Physics, SB RAS, Novosibirsk, 630090, Russia}
\affiliation{Novosibirsk State University, Novosibirsk, 630090, Russia}
\author{A.~P.~Kryukov}
\affiliation{Budker Institute of Nuclear Physics, SB RAS, Novosibirsk, 630090, Russia}
\author{N.~A.~Melnikova}
\affiliation{Budker Institute of Nuclear Physics, SB RAS, Novosibirsk, 630090, Russia}
\affiliation{Novosibirsk State University, Novosibirsk, 630090, Russia}
\author{N.~Yu.~Muchnoy}
\affiliation{Budker Institute of Nuclear Physics, SB RAS, Novosibirsk, 630090, Russia}
\affiliation{Novosibirsk State University, Novosibirsk, 630090, Russia}
\author{A.~E.~Obrazovsky}
\affiliation{Budker Institute of Nuclear Physics, SB RAS, Novosibirsk, 630090, Russia}
\author{E.~V.~Pakhtusova}
\affiliation{Budker Institute of Nuclear Physics, SB RAS, Novosibirsk, 630090, Russia}
\author{K.~V.~Pugachev}
\affiliation{Budker Institute of Nuclear Physics, SB RAS, Novosibirsk, 630090, Russia}
\affiliation{Novosibirsk State University, Novosibirsk, 630090, Russia}
\author{S.~A.~Rastigeev}
\affiliation{Budker Institute of Nuclear Physics, SB RAS, Novosibirsk, 630090, Russia}
\author{Yu.~A.~Rogovsky}
\affiliation{Budker Institute of Nuclear Physics, SB RAS, Novosibirsk, 630090, Russia}
\affiliation{Novosibirsk State University, Novosibirsk, 630090, Russia}
\author{A.~I.~Senchenko}
\affiliation{Budker Institute of Nuclear Physics, SB RAS, Novosibirsk, 630090, Russia}
\author{S.~I.~Serednyakov}
\affiliation{Budker Institute of Nuclear Physics, SB RAS, Novosibirsk, 630090, Russia}
\affiliation{Novosibirsk State University, Novosibirsk, 630090, Russia}
\author{Z.~K.~Silagadze}
\affiliation{Budker Institute of Nuclear Physics, SB RAS, Novosibirsk, 630090, Russia}
\affiliation{Novosibirsk State University, Novosibirsk, 630090, Russia}
\author{I.~K.~Surin}
\affiliation{Budker Institute of Nuclear Physics, SB RAS, Novosibirsk, 630090, Russia}
\author{Yu.~V.~Usov}
\affiliation{Budker Institute of Nuclear Physics, SB RAS, Novosibirsk, 630090, Russia}
\author{A.~G.~Kharlamov}
\affiliation{Budker Institute of Nuclear Physics, SB RAS, Novosibirsk, 630090, Russia}
\affiliation{Novosibirsk State University, Novosibirsk, 630090, Russia}
\author{D.E.Chistyakov}
\affiliation{Budker Institute of Nuclear Physics, SB RAS, Novosibirsk, 630090, Russia}
\affiliation{Novosibirsk State University, Novosibirsk, 630090, Russia}
\author{Yu.~M.~Shatunov}
\affiliation{Budker Institute of Nuclear Physics, SB RAS, Novosibirsk, 630090, Russia}
\affiliation{Novosibirsk State University, Novosibirsk, 630090, Russia}
\author{S.~P.~Sherstyuk}
\affiliation{Budker Institute of Nuclear Physics, SB RAS, Novosibirsk, 630090, Russia}
\affiliation{Novosibirsk State University, Novosibirsk, 630090, Russia}
\author{D.~A.~Shtol}
\affiliation{Budker Institute of Nuclear Physics, SB RAS, Novosibirsk, 630090, Russia}

\collaboration{SND Collaboration}

\begin{abstract}
The results of the search for the $e^+e^-\to\eta^{\prime}\gamma$ process in
the center-of-mass energy range from 1.075 to 2 GeV are
presented. We analyze data with an integral luminosity of 746 pb$^{-1}$
accumulated with the SND detector at the $e^+e^-$ collider
VEPP-2000 in 2010--2024. The $\eta^{\prime}\to\eta\pi^0\pi^0$ decay is 
used in the analysis, with the subsequent $\eta$ and $\pi^0$ decays into 
$\gamma\gamma$. Upper limits not exceeding 13 pb have been set on the
$e^+e^-\to\eta^{\prime}\gamma$ cross section at the 90\% confidence level.
\end{abstract}

\maketitle

\section{Introduction}
This work is devoted to the measurement of the $e^+e^- \to \eta^{\prime}\gamma$
cross section with the SND detector at the VEPP-2000 collider~\cite{VEPP2000}
in the center-of-mass energy region $1.075< \sqrt{s}<2$~GeV. The process 
$e^+e^- \to \eta^{\prime}\gamma$ is poorly studied experimentally. The vector
mesons decays $\phi$, $J/\psi$, and $\psi(2S)$ to 
$\eta^{\prime}\gamma$~\cite{pdg} and the cross section at
$\sqrt{s}=3.773$~GeV~\cite{cleoetap} and
10.6~GeV~\cite{babaretap} were measured. In the energy region under study,
there are the SND results~\cite{SNDetapgam2020} obtained using 12\% of the 
SND data collected to date. In the work~\cite{SNDetapgam2020} the upper limits
on the cross-section are set at the 90\% confidence level (CL): 
$\sigma_{\eta^{\prime}\gamma}<28$~pb at $1.15<\sqrt{s}<1.39$~GeV and
$\sigma_{\eta^{\prime}\gamma}<12$~pb at $1.39<\sqrt{s}<2.00$~GeV.

In the energy range from $\phi$ meson to 2 GeV, the $e^+e^-$ annihilation into
hadrons is dominated by contributions of excited vector resonances
of the $\rho$, $\omega$, and $\phi$ families. Therefore, we can expect that the
$e^+e^- \to \eta^{\prime}\gamma$ cross section will be determined by the 
radiative decays of these resonances. The quark model~\cite{qmodel} predicts
relatively large decay widths to $\eta^{\prime}\gamma$ for the
$\rho(1450)$ (60 keV) and $\phi(1680)$ (20 keV) states. Taking 
the $\rho(1450)$ and $\phi(1680)$ production cross section in $e^+e^-$ 
annihilation of 60 nb and 13 nb, respectively, we can estimate the 
corresponding cross sections 
$\sigma(e^+e^- \to \rho(1450) \to \eta^{\prime}\gamma)\approx 9$ pb 
and $\sigma(e^+e^- \to \phi(1680)\to \eta^{\prime}\gamma)\approx 2$ pb.

In this work, we expect to reach the sensitivity of the 
$e^+e^- \to \eta^{\prime}\gamma$ cross section measurement at the level of 
the quark model predictions.

\section{Detector and experiment}
To search for the process $e^+e^-\to\eta^{\prime}\gamma$, we use data with
an integral luminosity of about 746 pb$^{-1}$ collected
with the SND detector~\cite{SND} at the VEPP-2000 $e^+e^-$ collider from
2010 to 2024. The energy region 1.05--2.00 GeV was scanned several times
with a step of 20--25 MeV. 
In this analysis, due to the small value of the cross section, a small set
of statistics was obtained, and therefore we present as the result the cross
section values averaged over 4 energy intervals listed in the
table~\ref{tabl}.

A detailed description of the SND detector is given in Refs.~\cite{SND}.
This is a non-magnetic detector, the main part of which is a three-layer
spherical electromagnetic calorimeter based on NaI(Tl) crystals. The solid
angle of the calorimeter is 95\% of 4$\pi$. Its energy resolution for
photons is $\sigma_E/E=4.2\%/\sqrt[4]{E({\rm \mbox{GeV}})}$, and its angular
resolution is about $1.5^\circ$. The angles and production vertex of
charged particles are measured in a tracking system consisting of a
nine-layer drift chamber and a proportional chamber with 
cathode-strip readout. The track-system solid angle is 94\% of 4$\pi$.

The search for the $e^+e^-\to\eta^{\prime}\gamma$ process was discovered in
the $\eta^{\prime}\to\eta\pi^0\pi^0$ decay channel, with subsequent decays
$\eta$- and $\pi^0$-mesons in $\gamma\gamma$.
Since the final state for the
process under study does not contain charged particles, the process 
$e^+e^-\to\gamma\gamma$ is chosen for normalization.
As a result of this normalization, the systematic uncertainties associated
with the event selection in the first level trigger, as well as
the uncertainties arising due to superimposing of beam-generated background
charged tracks on events under study, are canceled. The accuracy of the
luminosity measurement using the $e^+e^-\to\gamma\gamma$ process is
2.2\%~\cite{SNDomegapi2013}.

\section{Event selection}
The process under study
\begin{equation}
e^+e^-\to\eta^{\prime}\gamma\to\eta\pi^0\pi^0\gamma\to 7\gamma
\label{etapg7g}
\end{equation}
contains seven photons in the final state.
Therefore, we select events with exactly seven reconstructed photons with
energy greater 20 MeV and no charged tracks.
The total energy deposition in the calorimeter $E_{\rm tot}$ and the 
total event momentum $P_{\rm tot}$ calculated using the energy depositions in
the calorimeter crystals are required to satisfy the following
conditions:
\begin{eqnarray}
&0.7 < E_{\rm tot}/\sqrt{s} < 1.2,\nonumber\\
&P_{\rm tot}/\sqrt{s} < 0.3,\nonumber\\
&E_{\rm tot}/\sqrt{s} - P_{\rm tot}/\sqrt{s} > 0.7.
\end{eqnarray}

For selected events, kinematic fits are performed using
measured photon angles and energies, energy-momentum conservation laws, and
assumptions about the presence of intermediate $\pi^0$ and $\eta$ mesons.
As a result of the fit, the photon energies are refined and
$\chi^2$ is calculated for the used kinematic hypothesis. Two hypotheses are
tested: 
\begin{itemize}
\item $e^+e^-\to 7\gamma$ ($\chi^2_{7\gamma}$),
\item $e^+e^-\to\eta\pi^0\pi^0\gamma\to 7\gamma$ ($\chi^2_{\eta\pi^0\pi^0\gamma}$).
\end{itemize}
Events with $\chi^2_{\eta\pi^0\pi^0\gamma} < 50$ are selected for
further analysis.

The main sources of background are the processes 
$e^+e^ -\to\omega\eta\pi^0$,
$e^+e^-\to\omega\pi^0\pi^0$,
$e^+e^-\to\eta\gamma$, 
$e^+e^-\to\pi^0\pi^0\gamma$, 
$e^+e^-\to\eta\pi^0\gamma$, and
$e^+e^-\to\eta\eta\gamma$ 
with the decays $\omega\to\pi^0\gamma$, $\eta\to 3\pi^0$ or $\gamma\gamma$.
Processes with neutral kaons in the final state
$e^+e^-\to K_SK_L(\gamma)$, 
$e^+e^-\to K_SK_L\pi^0$,
$e^+e^- \to K_SK_L\pi^0\pi^0$, and 
$e^+e^-\to K_SK_L\eta$ with the decay $K_S\to 2\pi^0$
also contribute to the background.
In processes with the $K_L$ meson, additional photons originate from
the $K_L$ nuclear interaction in the calorimeter or its decay. 
Also, additional photons arise from splitting of electromagnetic showers,
emission of photons by the initial particles at a large angle, and
superimposing of beam-generated background.

Photon parameters obtained after the kinematic fit in the 
$e^+e^-\to 7\gamma$ hypothesis are used to suppress the background. 
The events contening three pairs of photons with an invariant
mass satisfying the condition $ | M_{\gamma\gamma}-M_{\pi^0} |<35$~MeV
were rejected.
To suppress the background from processes containing the $\omega$-meson, all
possible three-photon combinations in an event are tested. If a
combination with the three-photon invariant mass 
$| M_{3\gamma}-M_{\omega} |<35$~MeV and the invariant mass of two of the three
photons $| M_{\gamma\gamma}-M_{\pi^0} |<35$~MeV is found, the event is 
rejected.

\begin{figure*}
\centering
\includegraphics[width=0.55\textwidth]{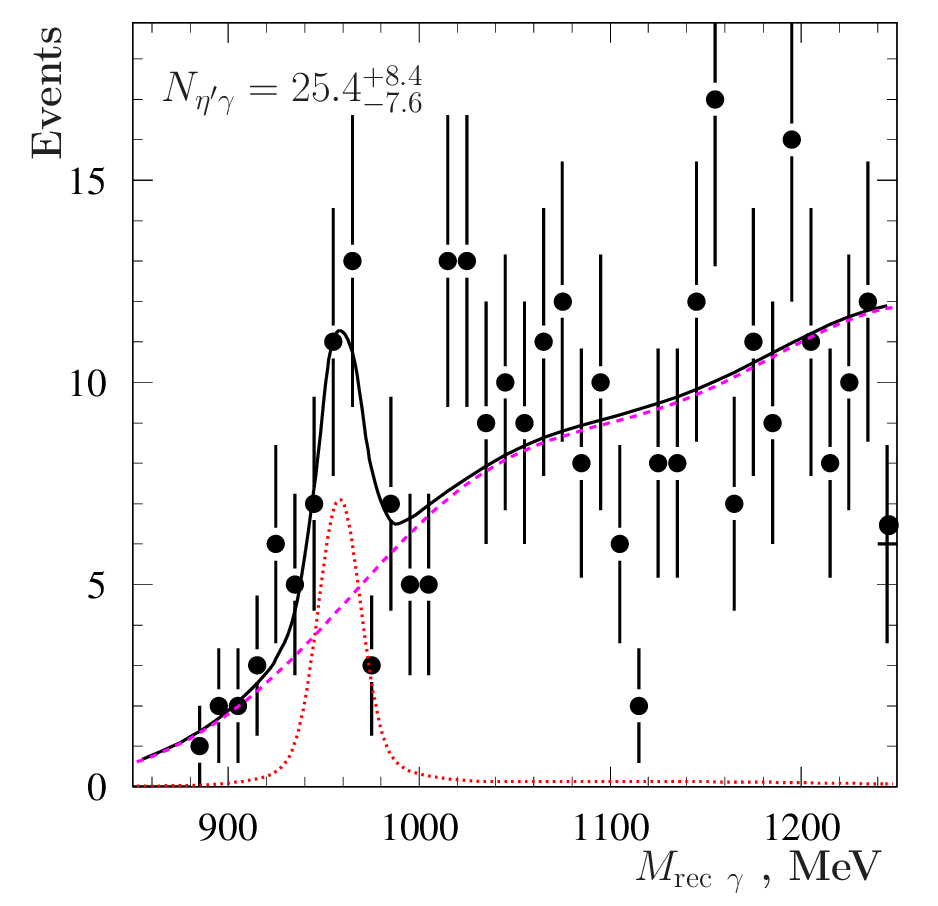}
\caption{The $M_{\rm rec}$ distribution for all selected data events (points with
error bars). The solid curve represents the result of the fit described
in the text. The dashed curve shows the fitted background contribution.
The dotted curve is the signal line shape.}
\label{fig1}
\end{figure*}

To determine the number of $\eta^{\prime}\gamma$ events, we analyze the
distribution over the invariant recoil mass against the photon $M_{\rm rec}$,
calculated after the kinematic fit in the hypothesis
$e^+e^-\to\eta\pi ^0\pi^0\gamma$. This distribution in the range $850 <
M_{\rm rec} < 1250$ MeV for all selected events is shown in Fig.~\ref{fig1}.

It is fitted by a sum of signal and background:
\begin{eqnarray}
P(M_{\rm rec}) = N_{\eta^{\prime}\gamma}P_{\eta^{\prime}\gamma}(M_{\rm rec}) +
N_{\rm bkg}P_{\rm bkg}(M_{\rm rec}),
\label{Ffit}
\end{eqnarray}
where the functions $P_{\eta^{\prime}\gamma}$ and $P_{\rm bkg}$ are
normalized to unity. The free fit parameters are the numbers of
signal ($N_{\eta^{\prime}\gamma}$) and background
($N_{\rm bkg}$) events. 

The signal line shape is obtained from simulation and then fitted by a 
sum of three Gaussian functions. The distribution for background is also
determined from simulation. The contributions of various background
processes are calculated using experimental data for the cross sections
for 
$e^+e^-\to\eta\gamma$~~\cite{SNDetagam2007},
$e^+e^-\to\pi^0\pi^0\gamma$~\cite{SNDpi0pi0g2016},
$e^+e^-\to\eta\pi^0\gamma$~\cite{SNDomegaeta2019,SNDetapi0g2020},
$e^+e^-\to\eta\eta\gamma$~\cite{SNDomegaeta2019,CMDkketa2019},
$e^+e^-\to \omega\pi^+\pi^-$~\cite{CMDomegapipi2000,BABARomegapipi2007},
$e^+e^-\to\omega\eta\pi^0$~\cite{SNDomegaetapi2016,BABARomegaetapi2018,BABARomegaetapi2021},
$e^+e^-\to K_SK_L(\gamma)$~\cite{BABARkskl2014},
$e^+e^-\to K_SK_L\pi^0$~\cite{BABARkkpi2017,SNDkkpi2018},
$e^+e^-\to K_SK_L\pi^0\pi^0$~\cite{BABARkkpi2017},
and $e^+e^-\to K_SK_L\eta$~\cite{BABARkkpi2017}.
For the process $e^+e^-\to \omega\pi^0\pi^0$, the isotopic relation
$\sigma(\omega\pi^+\pi^-)=2\sigma(\omega\pi^0\pi^0)$ is used. Upon
calculation the background, radiation corrections~\cite{radcor} are taken
into account. This is especially important for the processes
$e^+e^-\to K_S K_L(\gamma)$ and $e^+e^-\to\eta\gamma(\gamma)$, which are
dominated by ``radiative return'' to the $\phi$ meson, i.e. the processes
$e^+e^-\to\phi\gamma\to K_S K_L \gamma$ and
$e^+e^-\to\phi\gamma\to\eta\gamma\gamma$. In the energy region above 1.6
GeV, the cross sections of some background processes are known with poor
accuracy ($\sim$ 25\%). The expected distribution of the total background
obtained from simulation (\ref{Ffit}) is normalized to unity and fitted by a
sum of three Gaussian functions.

The result of the fit to the $M_{\rm rec}$ spectrum by Eq.~(\ref{Ffit})
is shown in Fig.~\ref{fig1}. It is seen that the simulation
reproduces the shape of the background reasonably well. The number of 
background events $290\pm 20$ found in the fit is consistent
with that expected from simulation $230\pm 50$. The number of 
$e^+e^-\to\eta^{\prime}\gamma$ events is equal to
$N_{\eta^{\prime}\gamma}=25\pm8$. The significance of signal observation
obtained from the difference between the likelihood functions of the 
fit described above and the fit with $N_{\eta^{\prime}\gamma}=0$ is found to
be $4\sigma$.

A similar fit is performed for the $M_{\rm rec}$ distributions 
in the four energy intervals. The obtained numbers of events of the signal
and background are listed in Table~\ref{tabl}.

\section{Detection efficiency}
\begin{figure*}
\centering
\includegraphics[width=0.95\textwidth]{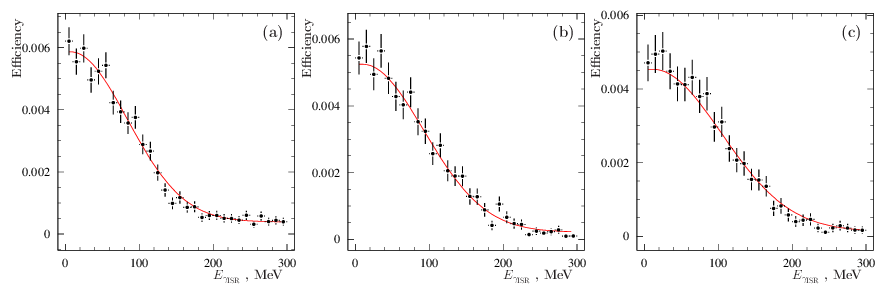}
\caption{The dependence of the detection efficiency for the
$e^+e^-\to\eta^{\prime}\gamma(\gamma)$ process on the energy of the
additional photon emitted by the initial particles at $\sqrt{s} = 1.5$ GeV
(a), 1.72 GeV (b), and 1.878 GeV (c). The points with error bars represent
the simulation distribution. The curve is the result of the fit 
by a smooth function.}
\label{fig2}
\end{figure*}

The detection efficiency of the process under study is determined using
Monte-Carlo simulation, which takes into account radiative corrections to the
initial state, in particular, the emission of additional photons. The
angular distribution of these photons is generated according to 
Ref.~\cite{BM}. Fig.~\ref{fig2} shows the dependence of the detection
efficiency $\varepsilon(\sqrt{s},E_{\gamma_{\rm ISR}})$ on the energy 
of the photon emitted from the initial state $E_{\gamma_{\rm ISR}}$ at 
three center-of-mass energies.

The values of the detection efficiency at $E_{\gamma_{\rm ISR}}=0$
averaged over the energy intervals are listed in Table~\ref{tabl}.

\section{Born cross section \label{fit}}
The visible cross section for the process $e^+e^-\to \eta^{\prime}\gamma$ 
directly measured in experiment 
is related to the Born cross section ($\sigma(\sqrt{s})$) by the 
following formula:
\begin{equation}
\label{viscrs}
\sigma_{\rm vis}(\sqrt{s}) = \int\limits_{0}^{x_{\rm max}}
\varepsilon \left( \sqrt{s},\frac{x\sqrt{s}}{2} \right)
F\left( x,\sqrt{s} \right)
\sigma \left( \sqrt{s(1-x)} \right) dx~,
\end{equation}
where $F(x,\sqrt{s})$ is a function describing the distribution over the energy
fraction $x=2E_{\gamma_{\rm ISR}}/\sqrt{s}$ carried away by
photons emitted from the initial state~\cite{radcor}, 
$x_{\rm max}=1-m^2_{{\eta}^{\prime}}/s$.
The expression (\ref{viscrs}) can be rewritten as:
\begin{equation}
\label{viscrs1}
\sigma_{\rm vis}(\sqrt{s}) =
\varepsilon_0(\sqrt{s})\,\sigma(\sqrt{s})\,(1+\delta(\sqrt{s}))~,
\end{equation}
where the detection efficiency $\varepsilon_0(\sqrt{s})$ and the radiative
correction $\delta(\sqrt{s})$ are defined as follows:
\begin{equation}
\varepsilon_0(\sqrt{s}) \equiv \varepsilon(\sqrt{s},0),
\end{equation}
\begin{equation}
\delta(\sqrt{s}) = \frac{\int\limits_{0}^{x_{\rm max}}
\varepsilon \left( \sqrt{s},\frac{x\sqrt{s}}{2} \right) ~F(x,\sqrt{s})~
\sigma \left( \sqrt{(1-x)s} \right) dx
}{\varepsilon(\sqrt{s},0)\sigma(\sqrt{s})}-1.
\label{rc}
\end{equation}
Technically, the Born cross section is determined as follows. The energy 
dependence of the measured visible cross section
$\sigma_{\rm vis}(\sqrt{s_i}) = N_{\eta^{\prime}\gamma,i}/IL_i$, where $i$ is
the energy interval number, is fitted with Eq.~(\ref{viscrs}). 
To parameterize the Born cross section, a theoretical model that describes
the data well was chosen.
With the obtained parameters of the
theoretical model, the radiation correction $\delta(\sqrt{s_i})$ is
determined by Eq.~(\ref{rc}), and then the experimental Born cross
section~$\sigma(\sqrt{s_i})$ is calculated by Eq.~(\ref{viscrs1}).

To approximate the cross section, a model with one effective resonance is
used:
\begin{equation}
\label{parcrs0}
\sigma_{\eta^{\prime}\gamma}(\sqrt{s}) = 
\left(\frac{k_\gamma(\sqrt{s})}{\sqrt{s}} \right)^3 \cdot \nonumber\\
\end{equation}
\begin{equation}
\label{parcrs1}
\cdot \left| \frac{m_V \Gamma_V(m_V)}{m^2_V - s - i \sqrt{s}\Gamma_V}
\sqrt{ \frac{m^3_V} {k_\gamma(m_V)^3} \sigma_{V\eta^{\prime}\gamma}}
\right|^2,
\end{equation}
\begin{equation}
\label{parcrs2}
k_\gamma(\sqrt{s}) = \frac{\sqrt{s}}{2} \left( 1 - \frac{m^2_{\eta^{\prime}}}{s} \right),
\end{equation}
where $m_V$ and
$\Gamma_V(\sqrt{s})$ are the resonance mass and total width,
$\sigma_{V\eta^{\prime}\gamma}$ is the cross section at the resonance maximum.
\begin{figure*}
\centering
\includegraphics[width=0.5\textwidth]{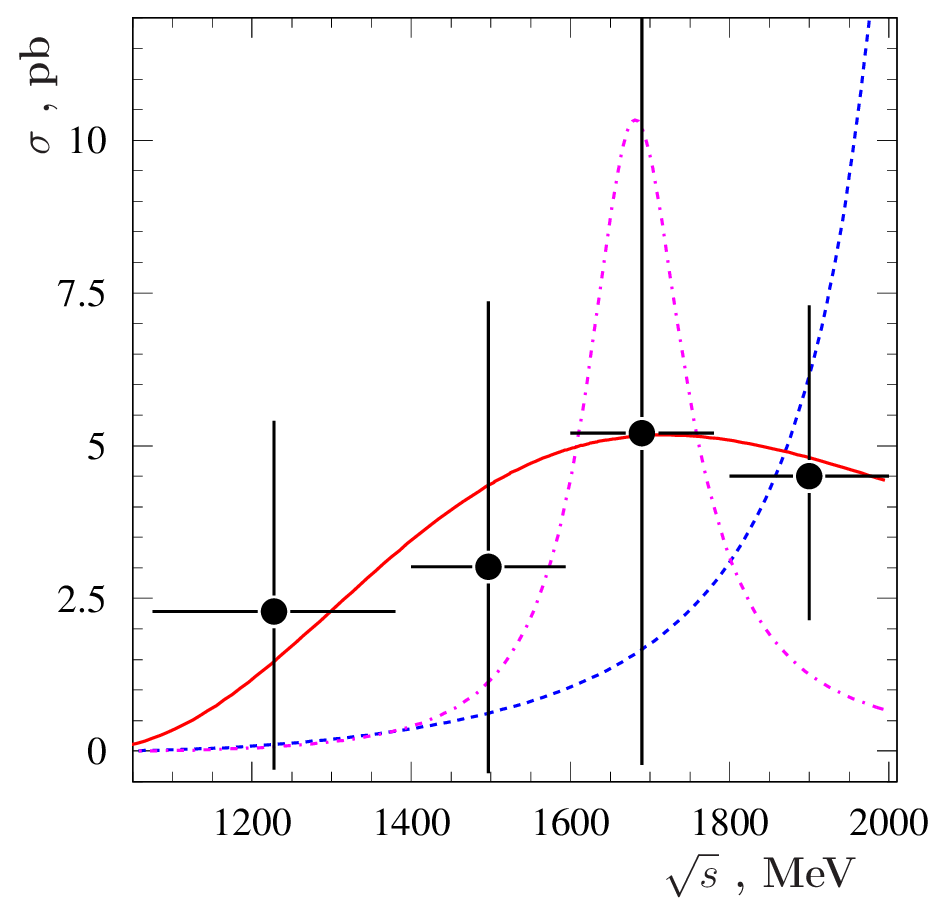}
\caption{The cross section for the process $e^+e^-\to\eta^{\prime}\gamma$ measured in 
this work (points with error bars). The solid curve is the result of
the fit with a resonance with free mass and width, the dashed and dash-dotted
curves are the results of the fits with $\phi(2170)$ and $\phi (1680)$,
respectively.}
\label{fig3}
\end{figure*}
The free fit parameters are $m_V$, $\Gamma_V$ and
$\sigma_{V\eta^{\prime}\gamma}$. The radiative corrections calculated in
this model are listed in Table~\ref{tabl}. The Born cross section values
obtained using Eq.~(\ref{viscrs1}) are listed in Table~\ref{tabl} and
shown in Fig.~\ref{fig3}. The fitting curve is shown in
Fig.~\ref{fig3} by a solid line.
\begin{table*}
\caption{Energy interval ($\sqrt{s}$), integrated luminosity ($IL$), 
number of selected $e^+e^-\to\eta^{\prime}\gamma$ events
($N_{\eta^{\prime}\gamma}$), number of background events in the
range $850 < M_{\rm rec} < 1250$ MeV ($N_{\rm bkg}$), obtained from the fit
to $M_{\rm rec}$ spectrum and calculated using simulation, detection
efficiency ($\varepsilon_0$), radiation correction ($1+\delta$), and Born
cross section for the process $e^+e^-\to\eta^{\prime }\gamma$ 
($\sigma_{\eta^{\prime}\gamma}$). The
first error in the cross section is statistical, the second is
systematic.
The upper limit on the cross section at the 90\% CL is given in
parentheses.}
\label{tabl}
\begin{ruledtabular}
\begin{tabular}{ccccccc}
$\sqrt{s}$, GeV &
$IL$, pb$^{-1}$ &
$N_{\eta^{\prime}\gamma}$ &
$N_{\rm bkg}$ &
$\epsilon_0$, \% &
$1+\delta$ &
$\sigma_{\eta^{\prime}\gamma}$, pb
\\ \hline \hline
1.075--1.38& 201 & $3.8^{+5.2}_{-4.3}$ & $ 69\pm 9 $ / $ 48 \pm  5$ & 0.89 & $0.87 \pm 0.01$ & $2.3^{+3.1}_{-2.6} \pm 0.1$ ( $<  5.8$ ) \\
1.4--1.594 & 120 & $2.5^{+3.6}_{-2.8}$ & $ 45\pm 7 $ / $ 26 \pm  6$ & 0.74 & $0.90 \pm 0.06$ & $2.9^{+4.1}_{-3.2} \pm 0.2$ ( $<  7.5$ ) \\
1.6--1.78  &  99 & $2.4^{+3.3}_{-2.5}$ & $ 39\pm 7 $ / $ 32 \pm  8$ & 0.60 & $0.90 \pm 0.01$ & $5.1^{+7.0}_{-5.3} \pm 0.2$ ( $< 13.1$ ) \\
1.8--2.0   & 326 & $8.2^{+5.1}_{-4.3}$ & $139\pm 12$ / $121 \pm 21$ & 0.54 & $0.91 \pm 0.16$ & $5.0^{+3.1}_{-2.7} \pm 1.0$ ( $<  8.6$ ) \\
\hline
\end{tabular}
\end{ruledtabular}
\end{table*}

To determine the model uncertainty of the radiation correction, the fits
are performed with $V=\phi (1680)$ and $\phi(2170)$. The masses and widths
of the resonances in this case are fixed at Particle Data Group 
values~\cite{pdg}. The fit results are shown in Fig.~\ref{fig3}. It is seen
that all three models do not contradict the data. The difference between the
radiative corrections calculated in the three models is used to estimate
the model uncertainty listed in Table~\ref{tabl}.

In Table~\ref{tabl}, the statistical and systematic uncertainties for the 
cross section are quoted. The latter includes uncertainties of  
the detection efficiency (3.3\%), luminosity measurements (2.2\%), and
radiation correction. A detailed study of the systematic uncertainties 
associated with the selection of multiphoton events was carried out in 
Refs.~\cite{SNDomegapi2013,SNDetagam2007,SNDpi0pi0g2016,SNDetagam2014}. This
uncertainty is estimated to be 3\%. 
Additionally it is nessecary to add  the uncertainty appering due to the
difference between data and simulation in the probability of photon
conversion before the tracking system equal to 1.3\%.

The obtained cross-section values in the first three energy intervals do not
contradict zero. The significance of signal observation in the last interval
is $2\sigma$. Therefore, the last column of Table~\ref{tabl} also
contains the upper limits on the cross section at the 90\% CL.

\section{Summary}
This paper presents the results of the search for the $e^+e^- \to
\eta^{\prime}\gamma$ process in the center-of-mass energy range from 1.07 
to 2.00 GeV. Data with an integrated luminosity of 746 pb$^{-1}$ accumulated
by the SND detector at the VEPP-2000 $e^+e^-$ collider in 2010--2024
have been analyzed. The decay mode $\eta^{\prime}\to\eta\pi^0\pi^0$
with the subsequent decays $\eta$ and $\pi^0\to\gamma\gamma$ has been used.

The significance of the $e^+e^-\to\eta^{\prime}\gamma$ signal in the full
selected data sample has been found to be $4\sigma$. 
However, after dividing the data into 4 energy intervals, there is no interval
with a significance greater than $2\sigma$. The upper limit on the cross 
section has been determined as not exceeding 13 pb at the 90\% CL. In the region of the 
$\rho(1450)$ resonance, the upper limit is 7.5 pb and is at the level 
of the quark model prediction~\cite{qmodel}.

\end{document}